\documentclass[12pt,a4paper]{article}
\pdfoutput=1
\usepackage{jheppub}
\usepackage{hyperref, amsmath, amssymb, graphicx, amsfonts, latexsym, bbold, mathbbol}
\usepackage{mathrsfs} 
\usepackage{tabstackengine}
\usepackage{color}
\newcommand{    \be}{\begin{equation}}
\newcommand{\ee}{\end{equation}}
\def\ccr{\nonumber\\} 
\newcommand{\DDslash}{D \! \! \! \!  / \, } 
\newcommand{\Aslash}{A \! \! \!  /} 
\newcommand{\Bslash}{B \! \! \! \! / \, } 
\newcommand{\bpsi}{\overline{\psi}}

\def\la{\langle}
\def\ra{\rangle}
\def\tr{{\rm tr}\,}

\title{Weyl fermions in a non-abelian gauge background and trace anomalies}

\author[a,b]{Fiorenzo Bastianelli}
\author{and}
\author[b]{Matteo Broccoli}

\affiliation[a] {Dipartimento di Fisica ed Astronomia, Universit{\`a} di Bologna and\\
INFN, Sezione di Bologna, via Irnerio 46, I-40126 Bologna, Italy}
\affiliation[b]{Max-Planck-Institut f\"ur Gravitationsphysik (Albert-Einstein-Institut)\\
 Am M\"uhlenberg 1, D-14476 Golm, Germany}
 \emailAdd{bastianelli@bo.infn.it} 
 \emailAdd{ matteo.broccoli@aei.mpg.de}
  
\abstract{We study the trace and chiral anomalies of Weyl fermions in a non-abelian gauge background in four dimensions.
Using a Pauli-Villars regularization we  identify the trace anomaly,  proving that it can be cast in a gauge invariant form, 
even in the presence of the non-abelian chiral anomaly, 
that we rederive to check the consistency of our methods. In particular, we find that the trace anomaly does not contain
any parity-odd topological contribution, whose presence has been debated in the recent literature.
}

\begin{document}
\maketitle
\flushbottom

\section{Introduction}

In this paper we calculate the trace (and chiral) anomalies of Weyl fermions coupled to non-abelian gauge fields in four dimensions.
One of the motivations to study this problem arises from 
a debate on  whether a topological, parity-odd term is present in the trace anomaly of the stress tensor of chiral fermions. 
We find that it does not.

We start by considering  Bardeen's method \cite{Bardeen:1969md} 
that embeds the Weyl theory into the theory of Dirac fermions coupled to vector 
and axial non-abelian gauge fields. Using a Pauli-Villars (PV) regularization we calculate its trace anomaly. 
As an aside we rederive the well-known non-abelian chiral  anomaly to check the consistency of our methods.
A chiral limit produces the searched for anomalies of the non-abelian Weyl fermions. 

\section{Bardeen's model}

We consider  the Bardeen's model of massless Dirac fermions $\psi$ coupled to vector and axial  
non-abelian gauge fields, $A_a$ and $B_a$. The lagrangian is given by
\be
{\cal L} = -\bpsi \DDslash(A,B)\psi  
\label{lag}
\ee
where $\DDslash(A,B)= \gamma^a D_a(A,B) $, with $D_a(A) =\partial_a + A_a+ B_a\gamma^5$  the covariant derivative 
for the gauge group $G\times G$. Taking an appropriate limit on the background (by setting $A_a=B_a \to \frac{A_a}{2}$) 
one finds the theory of left-handed Weyl fermions.
We  expand the gauge fields on the generators of the gauge group as $A_a= -i A_a^\alpha T^\alpha$ 
and $B_a= -i B_a^\alpha T^\alpha$. The components  $A_a^\alpha$ and $B_a^\alpha$ are real, 
and $T^\alpha$ denote the hermitian generators in the representation of $G$ chosen for $\psi$ 
(we allow for the presence of an abelian subgroup, for example 
one could consider the group $U(N)$ with the fermion $\psi$ sitting in the fundamental
representation)\footnote{The  generators satisfy the Lie algebra $ [T^\alpha, T^\beta] = i f^{\alpha \beta \gamma}T^\gamma$.
Our conventions for Weyl and Dirac fermions follow those made explicit in ref. \cite{Bastianelli:2018osv}.}.

This model is classically gauge invariant and conformally invariant. We wish to compute systematically the anomalies.
The chiral anomaly is well-known, of course,  and we recompute it to test our methods. The main aim is to 
obtain the trace anomaly. 

Let us first review the classical symmetries. The lagrangian is invariant under the $G\times G$ gauge transformations. 
Using infinitesimal parameters $\alpha=-i \alpha_a^\alpha T^\alpha $ and $\beta=-i \beta_a^\alpha T^\alpha $, they read
\be 
\left\{
\begin{aligned}
 & \delta \psi  = - ( \alpha + \beta \gamma^5 )\psi   \\
 & \delta \bpsi  = \bpsi (\alpha -  \beta \gamma^5) \\
 & \delta \psi_c  =   (\alpha^T  -\beta^T \gamma^5) \psi_c \\
 & \delta  A_a  =       \partial_a  \alpha + [A_a, \alpha] +[B_a, \beta]   \\
 &  \delta  B_a  =   \partial_a  \beta + [A_a, \beta]  +  [B_a, \alpha] 
\end{aligned}
\right.
\label{gauge-tra}
\ee
where $\psi_c = C^{-1} \bpsi^T$ is the charge conjugated spinor.
The transformations of the gauge fields can be written more compactly as 
\be
\delta {\cal A}_a  =   \partial_a  \tilde \alpha + [{\cal A}_a, \tilde \alpha] 
\ee
where ${\cal A}_a = A_a +B_a\gamma^5$ and $ \tilde \alpha = \alpha +\beta \gamma^5$.
The corresponding field strength 
\be
\mathscr{F}_{ab} = \partial_a  {\cal A}_b -\partial_b  {\cal A}_a  +  [{\cal A}_a , {\cal A}_b] = \hat F_{ab} + \hat G_{ab}\gamma^5 
\ee
contains  the Bardeen curvatures  $\hat F_{ab}$ and $\hat G_{ab}$ 
\be
\begin{aligned}
\hat F_{ab} &=  \partial_a  A_b -\partial_b  A_a  +  [A_a , A_b] +   [B_a , B_b] \\
\hat G_{ab} &=  \partial_a  B_b -\partial_b  B_a  +  [A_a , B_b]  +  [B_a , A_b] \;.
 \end{aligned}
 \label{Bardeen-curvatures}
\ee 
In the following we prefer to use the more explicit notation with $\gamma^5$.

One can use $A_a^\alpha$ and $B_a^\alpha$  as sources for the vector 
$J^{a\alpha}= i \bpsi \gamma^a T^\alpha \psi$ and axial $J_5^{a\alpha}= i \bpsi \gamma^a \gamma^5 T^\alpha\psi$
currents, respectively.
These currents are covariantly conserved on-shell, with the conservation law reading  
\be
\begin{aligned}
(D_a J^a)^\alpha\equiv 
\partial_a J^{a\alpha} - i \bpsi [\Aslash+\Bslash \gamma^5, T^\alpha] \psi =0 \\
(D_a J^a_5)^\alpha\equiv 
\partial_a J_5^{a\alpha} - i \bpsi [\Aslash \gamma^5+\Bslash, T^\alpha] \psi =0 
\end{aligned}
\ee
or, equivalently, as 
\be
\begin{aligned}
& (D_a J^a)^\alpha = 
\partial_a J^{a\alpha} + f^{\alpha\beta \gamma} A_a^\beta J^{a\gamma} + f^{\alpha\beta \gamma} B_a^\beta J_5^{a\gamma} =0 \\
& (D_a J^a_5)^\alpha=  
\partial_a J_5^{a\alpha} + f^{\alpha\beta \gamma} A_a^\beta J_5^{a\gamma} + f^{\alpha\beta \gamma} B_a^\beta J^{a\gamma} =0 \;.
\end{aligned}
\ee
Indeed, under an infinitesimal gauge variation of the external sources  $A_a^\alpha$ and $B_a^\alpha$, 
the action $S =\int d^4 x\, {\cal L}$ varies as 
\be
 \delta^{(A,B)} S = - \int d^4 x \,( \alpha^\alpha  (D_a J^a)^\alpha +   \beta^\alpha  (D_a J_5^a)^\alpha )   \;,
 \label{uno}
\ee
and the gauge symmetries implies that $J^{a\alpha}$ and $J_5^{a\alpha}$ are covariantly conserved on-shell, as stated above.

Similarly, to study the stress tensor, it is useful to couple the theory to gravity by introducing the vierbein $e_\mu{}^a$  and related 
spin connection. One may verify that the model acquires a Weyl invariance, i.e. an invariance under 
arbitrary local scalings of the vierbein. 
This suffices to prove conformal invariance in flat space. The vierbein is used also as an external source for the stress tensor
\be
T^{\mu a} = \frac{1}{e}\frac{\delta S}{\delta e_{\mu a}}
\ee
where $e$ denotes the determinant of the vierbein. The Weyl symmetry implies that the stress tensor is traceless on-shell.
Indeed, an infinitesimal  Weyl transformation with local parameter $\sigma $ is of the form
\be 
	\left\{
	\begin{aligned}
   &\delta \psi  = - \frac{3}{2}\sigma  \psi   \\
     &\delta  A_a  =  \delta  B_a  =  0\\
      &\delta e_\mu{}^a  =  \sigma e_\mu{}^a 
		\end{aligned}
	\right.
	\label{weyl-tra}
\ee
and  varying the action under an infinitesimal Weyl transformation of the vierbein only produces the trace of the stress tensor
\be
\delta^{(e)} S = \int d^4 x e \, \sigma   T^a{}_a \;.
\label{due}
\ee
Then, the full Weyl symmetry implies that the trace vanishes on-shell, $T^a{}_a =0$.

\section{PV regularization} 

To regulate the one-loop effective action we introduces massive PV fields.  The mass term produces the anomalies, 
which we will compute with heat kernel methods. 

We denote by $\psi$ the PV fields as well (for the moment this does not cause any confusion)
and add  a Dirac mass term to their massless lagrangian  in \eqref{lag} 
\be
\Delta {\cal L} = - M \bpsi  \psi =   \frac{M}{2} (\psi^T_c C \psi +\psi^T C \psi_c) \;.
\label{mass}
\ee
It preserves  vector gauge invariance but breaks axial  gauge invariance. Indeed under \eqref{gauge-tra}
the mass term varies as
\be
\delta \Delta {\cal L} = 2 M \bpsi \beta \gamma^5 \psi = - M  (\psi^T_c \beta C \gamma^5 \psi 
+\psi^T \beta^T C \gamma^5 \psi_c)
\ee
where $\beta =-i\beta^\alpha T^\alpha$,
which shows that the vector gauge symmetry is preserved, leaving room for an  anomaly in the axial gauge symmetry.

The mass term sources also a trace anomaly, as the curved space version of \eqref{mass}
\be
\Delta {\cal L} = - e M \bpsi  \psi =   \frac{e M}{2} (\psi^T_c C \psi +\psi^T C \psi_c)
\label{cov-mass}
\ee
varies under the infinitesimal Weyl transformation \eqref{weyl-tra} as
 \be
\delta \Delta {\cal L} = - e \sigma M \bpsi  \psi =   \frac{e \sigma M}{2} (\psi^T_c C \psi +\psi^T C \psi_c) \;.
\ee
However, it preserves the general coordinate and local Lorentz symmetries.
One concludes that only axial gauge and trace anomalies are to be expected.

Casting the PV lagrangian   ${\cal L}_{\scriptscriptstyle PV} = {\cal L} + \Delta{\cal L}$
in the form 
\be 
{\cal L}_{\scriptscriptstyle PV} =  \frac12 \phi^T T  {\cal O} \phi +\frac12 M \phi^T T \phi 
\label{PV-lag}
\ee 
where  $ \phi  = \left( \begin{array}{c}    \psi \\  \psi_c    \end{array} \right)  $, allows us to recognize the operators 
\be
T {\cal O} =
\left( \begin{array}{cc}
  0 &  C \DDslash(-A^T,B^T)    \\   C \DDslash(A,B)  &0  
\end{array} \right)  \;, \qquad 
T  = \left( \begin{array}{cc}    0& C  \\  C &  0    \end{array} \right)  
\ee
and 
\be
{\cal O} =
\left( \begin{array}{cc}
  \DDslash(A,B)   &0  \\  0&  \DDslash(-A^T,B^T)  
\end{array} \right)  \;, \qquad 
{\cal O}^2 =
\left( \begin{array}{cc}
  \DDslash^2(A,B)   &0  \\  0 &  \DDslash^2(-A^T,B^T) 
\end{array} \right)  \;.
\label{PVD-Dmass}
\ee
The latter identifies the regulators, as we shall see in the next section.

\section{Regulators and consistent anomalies}

Using the Pauli-Villars regularization, we relate the anomaly computation to a sum of heat kernel traces,  
following the scheme of refs. \cite{Diaz:1989nx, Hatsuda:1989qy} which we briefly review.
Starting with a lagrangian for $\varphi$ 
 \be 
{\cal L} = \frac12 \varphi^T T {\cal O} \varphi 
\label{lagrangian}
\ee
invariant under a linear symmetry 
\be
\delta \varphi = K\varphi 
\ee
acting also on the backgroud fields contained in the
operator  $T {\cal O}$, one constructs  the one-loop effective action $\Gamma$ by a path integral.
The latter is regulated by subtracting a loop  of a massive PV field $\phi$ with lagrangian  \eqref{PV-lag}
\be
e^{i \Gamma} =\int D\varphi\; e^{i S}  \qquad \to \qquad e^{i \Gamma} =
\int D\varphi D\phi\;  e^{i (S + S_{\scriptscriptstyle PV})} 
\ee
where it is understood that one should take the decoupling limit $M\to \infty$, with all divergences canceled by renomalization.
The anomalous response of the path integral under a symmetry is due to the PV mass term only, 
as one can define the measure of the PV field to make the whole path integral  measure invariant.
In a hypercondensed notation, where  a term like $\phi^T \phi$
includes in the sum of the (suppressed) indices a  spacetime integration as well, the lagrangian in \eqref{lagrangian}
is equivalent to the  action, and one computes the symmetry variation of the regulated path integral as follows
\begin{align}
i\delta \Gamma & =i \la\delta S\ra =  \lim_{M \to \infty} \ i M \la \phi^T (TK +\frac12 \delta T) \phi \ra
\ccr
&= - \lim_{M \to \infty}  
{\rm Tr} \biggl [\biggl (K + \frac12 T^{-1} \delta T \biggr ) \biggl ( 
1+ \frac{\cal O}{M} \biggr )^{\!\!\! -1} \biggr ] 
\end{align}
where brackets $\la ...\ra$ indicate normalized correlators. It is convenient to manipulate this expression further, 
by using the identity $ 1 = (1 - \frac{\cal O}{M}) (1 - \frac{\cal O}{M})^{-1}$ 
and invariance of  the massless action, to cast it in the equivalent form 
 \be
i\delta \Gamma =i \la\delta S\ra  =
- \lim_{M \to \infty}  
{\rm Tr} \biggl [\biggl (K + \frac12 T^{-1} \delta T + \frac12 \frac{\delta {\cal O}}{M} \biggr ) 
\biggl ( 1- \frac{{\cal O}^2}{M^2} \biggr )^{\!\!\! -1} \biggr ]  \;.
\label{3.8}
\ee
In the derivation  we have considered a fermionic theory, used the PV propagator
\be
\la \phi \phi^T\ra = \frac{i}{T {\cal O} + T M} \;,
\ee
taken into account the opposite sign for the PV loop, and considered an invertible matrix $T$. 
In the limit $M\to \infty$ the regulating factor $ ( 1- \frac{{\cal O}^2}{M^2})^{-1}$ in \eqref{3.8}
can be effectively replaced by $e^{ \frac{{\cal O}^2}{M^2}}$, if ${\cal O}^2$ is negatively defined
(in euclidean).
This substitution allows us to use 
well-known heat kernel formulae.
Obviously, a symmetry remains anomaly free if one finds a symmetrical mass term.  

Thus, denoting 
\be
J=K + \frac12 T^{-1} \delta T + \frac12 \frac{\delta {\cal O}}{M} \;, \qquad { R}=-{\cal O}^2
\label{jac}
\ee
the anomaly is related to the trace of the heat kernel  of the regulator  ${R}$ with insertion of the operator $J$
\be
i\delta \Gamma =i \la\delta S\ra  =
- \lim_{M \to \infty}   {\rm Tr} [J e^{ -\frac{R}{M^2}}] \;.
\label{tra}
\ee
It has the same form of the regulated Fujikawa's trace producing the anomalies \cite{Fujikawa:1979ay, Fujikawa:1980vr},
where $J$ is the infinitesimal part of the jacobian arising from a change of  variables in the path integral
under a symmetry transformation, and $R$ is the regulator. 
The limit extracts only the mass independent term (negative powers of the mass vanish in the limit, while positive 
powers are renormalized away, usually by employing additional PV fields).
The PV method guarantees that the regulator $R$ together with the jacobian $J$ produces consistent anomalies,
which follows from the fact that one is computing directly the variation of the effective action.
 
Let us now go back to the specific case of the Bardeen's model,  and extract the  heat kernel traces that compute the anomalies.
For each symmetry we must consider the transformation generated by $K$ and obtain the corresponding form of $J$.

To start with, the vector current $J^{a\alpha}$ remains covariantly conserved also at the quantum level,
as the PV mass term is invariant under vector gauge transformations.

For the axial current, recalling the transformation laws in \eqref{gauge-tra}, one finds 
\be
J= \left( \begin{array}{cc}   i\beta^\alpha T^\alpha \gamma^5 
& 0  \\  0&  i\beta^\alpha T^{\alpha T} \gamma^5       \end{array} \right)  
\ee
as $\delta T$ vanishes,  while the contribution  from $\delta {\cal O}$ is also seen to vanish 
(all possible terms vanish under the Dirac trace). 
Here, $T^{\alpha T} $ denotes the  transposed of  $T^{\alpha} $. 
Removing the spacetime integration and the local parameters $\beta^\alpha$ from \eqref{tra}, 
and recalling the nomalizations in \eqref{uno}, \eqref{tra} and \eqref{hkt}, one finds
\be
(D_a \la J^a_5 \ra)^\alpha
=   
\frac{i}{(4\pi)^2} \Big [\tr [\gamma^5 T^\alpha a_{2} ({R}_\psi)] +\tr [\gamma^5 T^{\alpha T} a_{2} ({R}_{\psi_c}) ] \Big]
\label{Bardeen-anomaly}
\ee
where the remaining trace is the finite dimensional one on the gamma matrices and color space.
Here, we  find the  so-called Seeley-DeWitt coefficients $a_2({R_i})$ corresponding to the regulators $R_i$ 
associated to  the  fields assembled into $\phi$
\be
{R}_\psi   = -  \DDslash^2(A,B) \;,
\qquad
{R}_{\psi_c} = - \DDslash^2(-A^T,B^T) \;.
\label{regu}
\ee
 The  $a_2$  coefficients are the only ones that survive renormalization and the limit ${M \to \infty}$.

Similarly, for the Weyl symmetry one uses  the transformations in \eqref{weyl-tra} to find
\be
\la T^a{}_a \ra = -\frac{1}{2 (4\pi)^2}  \Big [ \tr a_{2}({R}_\psi)  +\tr a_{2}({R}_{\psi_c}) \Big ] 
\label{trace-anomaly}
\ee
where now also $\delta T$ contributes to \eqref{jac}, while $\delta {\cal O}$  drops out as before.
Again, all remaining traces are in spinor and color spaces. 
Since the mass term is general coordinate and local Lorentz invariant, no anomalies arise in those symmetries.

\section{Anomalies}

We are left to compute the anomalies produced by the traces of the heat kernel coefficients  $a_2$ in 
\eqref{Bardeen-anomaly} and \eqref{trace-anomaly}, with the  regulators \eqref{regu}. 
The heat kernel formulae needed in the calculation are well-known, and 
for commodity we have reported them in appendix \ref{appA}.  

The vector symmetry is guaranteed to remain anomaly free by the invariance of the mass term. As a check  
one may verify, using the explicit traces given in appendix \ref{appA}, that the would-be anomaly vanishes 
\be
(D_a \la J^a\ra)^\alpha
=   
\frac{i}{(4\pi)^2} \Big [\tr [ T^\alpha a_{2} ({R}_\psi)]  -  \tr [ T^{\alpha T} a_{2} ({R}_{\psi_c}) ] \Big] =0 \;.
\ee

\subsection{Chiral anomaly} 

Evaluation of \eqref{Bardeen-anomaly} produces the chiral anomaly 
\be
\begin{aligned}
(D_a \la J^a_5 \ra)^\alpha &=
 -\frac{1}{(4\pi)^2} \epsilon^{abcd}\, \tr_{_{\!\! YM}}  T^\alpha
\bigg [  \hat F_{ab} \hat F_{cd}  + \frac13 \hat G_{ab} \hat G_{cd}   \\
&-\frac83 (\hat F_{ab} B_c B_d + B_a \hat F_{bc} B_d +  B_a B_b\hat F_{cd})
+\frac{32}{3} B_a B_b B_c B_d  \bigg ]  \\
&+ PETs
\end{aligned}
\label{axial-anomaly2}
\ee
where the remaining trace is only in color space (the trace on gamma matrices has been implemented).
$PETs$ indicate the parity-even terms that take the form
\be
\begin{aligned}
PETs  &=
\frac{i}{(4\pi)^2} \tr_{_{\!\! YM}}  T^\alpha
\bigg [ \frac43 D^2 D B +
\frac23 [\hat F^{ab},\hat G_{ab}] +\frac83 [D^a \hat F_{ab}, B^b]  
\\
&- \frac43 \{B^2, DB \} + 8 B_a DB B^a +\frac83 \{ \{ B^a,B^b\}, D_a B_b \}
\bigg ]  \;.
\end{aligned}
\label{pet}
\ee
They are canceled by the chiral gauge variation of a local counterterm
\be
\begin{aligned}
\Gamma_{ct} = \int \frac{d^4x}{(4\pi)^2} \, \tr_{_{\!\! YM}} 
&\bigg [ \frac23 (D^aB^b)( D_a B_b)  +4F^{ab}(A)B_aB_b
- \frac83 B^4  +\frac43 B^a  B^b  B_a  B_b  \bigg ]
\end{aligned}
\label{ct1}
\ee
and the remaining answer coincides with the famous result obtained by Bardeen \cite{Bardeen:1969md}.

\subsection{Trace anomaly}
Evaluation of \eqref{trace-anomaly} produces the trace anomaly 
\be
\begin{aligned}
\la T^a{}_a \ra &= \frac{1}{(4\pi)^2}  \tr_{_{\!\! YM}} 
 \bigg [ \frac23 \hat F^{ab} \hat F_{ab} +\frac23 \hat G^{ab} \hat G_{ab} \bigg]  
+ CTTs 
\end{aligned}
\label{trace-anomaly2}
\ee
where $CTTs$ are the cohomologically trivial terms
\be
 CTTs = \frac{1}{(4\pi)^2} \bigg ( -\frac43  \bigg )
 \tr_{_{\!\! YM}}  \bigg [ D^2 B^2 +DBDB - (D^aB^b) (D_bB_a) - 2  F^{ab}(A) B_a B_b \bigg] 
\ee
 that are canceled by the  Weyl variation of the following counterterm 
\be
\begin{aligned}
\bar \Gamma_{ct} = \int \frac{d^4x \sqrt{g}}{(4\pi)^2} \, \tr_{_{\!\! YM}} 
&\bigg [ \frac23 (D^\mu B^\nu)( D_\mu B_\nu)  +4F^{\mu\nu}(A)B_\mu B_\nu +\frac13 R B^2
\bigg ]
\end{aligned}
\label{ct2}
\ee
where $\mu, \nu$ are curved indices, and $R$ the Ricci scalar. Of course, one restricts to flat space after variation.

The counterterms \eqref{ct1} and  \eqref{ct2} merge  
consistently into a unique counterterm that in curved space reads
\be
\begin{aligned}
\Gamma^{tot}_{ct} = \int \frac{d^4x \sqrt{g}}{(4\pi)^2} \, \tr_{_{\!\! YM}} 
&\bigg [ \frac23 (D^\mu B^\nu)^2  +4F^{\mu\nu}(A)B_\mu B_\nu +\frac13 R B^2
- \frac83 B^4  +\frac43 B^\mu  B^\nu  B_\mu  B_\nu  
\bigg ]\;.
\end{aligned}
\ee

One  may already notice that, on top of the complete gauge invariance of the trace anomaly,
there is no parity-odd term present. 

\section{Chiral and trace anomalies of Weyl fermions}

We are now ready to study the chiral limit of the Bardeen's model, and identify the chiral and trace anomalies
of Weyl fermions.
We take the limit $A_a=B_a \to \frac12 A_a$, which creates a chiral projector in the coupling to the gauge field, 
normalized as usual after the scaling. Then, $\hat F_{ab} = \hat G_{ab} \to  \frac12 F_{ab}(A)$ 
and $J^a=J^a_5 \to J_a= \frac12(J^a+J^a_5) $, so that from  \eqref{axial-anomaly2} and \eqref{trace-anomaly2} 
(without the cohomologically trivial terms)  we immediately derive the searched for anomalies for the  
left-handed Weyl fermions  coupled to non-abelian gauge fields
\be
\begin{aligned}
(D_a \la J^a \ra)^\alpha &=
-\frac{1}{(4\pi)^2} \epsilon^{abcd} \, \tr_{_{\!\! YM}}  T^\alpha \partial_a 
\bigg [ \frac23 A_b\partial_c  A_d  + \frac13 A_b A_c  A_d  \bigg ]  
\\
\la T^a{}_a \ra &= \frac{1}{(4\pi)^2}  \tr_{_{\!\! YM}} 
 \bigg [ \frac13  F^{ab}  F_{ab} \bigg ]  \;.
\end{aligned}
\ee

The chiral anomaly is the standard one, rederived as a check on the methods used here.
The trace anomaly is our new result, that verifies 
the absence of parity-odd terms. It is just half the trace anomaly of non-chiral Dirac fermions.

\section{Conclusions}

We have calculated the chiral and trace anomaly in the Bardeen's model of Dirac  fermions 
coupled to  non-abelian vector and axial
gauge fields, rederiving the famous result for the chiral anomaly and finding the trace anomaly.
Then, by a chiral limit  we have obtained the chiral and trace anomalies for 
left-handed Weyl fermions  coupled to non-abelian gauge fields.

The main aim of this paper was to find the explicit form of the trace anomaly for Weyl fermions, 
verifying that it does not contain any parity-odd term proportional to the topological  
Chern-Pontryagin density. The latter was conjectured to be a possibility in \cite{Nakayama:2012gu}, 
see also comments in \cite{Nakayama:2018dig, Nakayama:2019mpz}.
However, it was found to be absent in the abelian gauge coupling of a single Weyl fermion \cite{Bastianelli:2018osv}. 
Here we prove that it is absent also in the more general case of the coupling to non-abelian gauge fields.
The analogous case of a Weyl fermion on a curved spacetime background 
has been debated more extensively in the literature,
where  a topological term proportional to the  Pontryagin density had been reported in 
 \cite{Bonora:2014qla}, and confirmed  in \cite{Bonora:2017gzz, Bonora:2018obr},
where the concept of a MAT background, that extends the Bardeen construction to curved space, has been 
developed. However, the topological term was found to be absent in  
\cite{Bastianelli:2016nuf},  as confirmed also in \cite{Frob:2019dgf}.
We believe that the latter are the correct results. This conclusion indeed finds support from the 
analogous situation studied in this paper. Also, 
an analysis of a  Dirac fermion on the MAT background, suitably regularized with PV fields, does not seem to
produce parity-odd terms in the trace anomaly \cite{BB}. 

\acknowledgments{We wish to thank Loriano Bonora for stimulating discussions.}

\appendix 

\section{The heat kernel}
\label{appA} 

Let us consider  a flat $D$-dimensional spacetime  and an operator $H$ of the form  
\be
H= -\nabla^2 + V
\label{B1}
\ee
where $V$ is a matrix potential and $\nabla^2=\nabla^a \nabla_a$, with $\nabla_a=\partial_a + W_a$  
the gauge covariant derivative  satisfying
\be
[\nabla_a,\nabla_b] = \partial_a W_b-\partial_b W_a + [W_a, W_b] ={\cal F}_{ab}  
\label{B2}
\;.
\ee
The trace of the corresponding heat kernel has a  small time expansion given by
\be
\begin{aligned}
{\rm Tr} \left [ J  e^{-i s H} \right ] 
&=
\int d^Dx \, \tr  \left  [J(x) \la x|e^{-i s H} | x \ra \right] \\
&= \int \frac{d^Dx\, i}{(4 \pi i s)^{\frac{D}{2}}} \sum_{n=0}^\infty \tr [ J(x)a_n(x,H)] (is)^n
\\\
& =\int 
\frac{d^Dx\, i}{(4 \pi i s)^\frac{D}{2}}\, {\rm tr}\, [J(x)  (a_0(x,H)+ a_1(x,H) i s + a_2(x,H) (is)^2+...)] 
\label{hkt}
\end{aligned}
\ee
where the symbol ``tr'' is a trace on the remaining discrete matrix indices, 
$J(x)$ is an arbitrary matrix function,
and $a_n(x,H)$ are the heat kernel, or Seeley-DeWitt, coefficients.  
They are matrix valued, and the first ones are 
\be
\begin{aligned}
a_0(x,H) &= {\mathbb 1} \\
a_1(x,H) &=   - V \\
a_2(x,H) &= \frac12 V^2  -\frac{1}{6} \nabla^2 V + \frac{1}{12} {\cal F}_{ab}^2  
\end{aligned} 
\label{SdW}
\ee
where $\nabla_a V= \partial_a V+ [W_a, V]$, and so on.
More details on the heat kernel expansion can be found in  \cite{DeWitt:1965jb, DeWitt:1985bc}.
They have been computed with quantum mechanical path integrals in \cite{Bastianelli:1991be, Bastianelli:1992ct}, while
a useful report is \cite{Vassilevich:2003xt}.

In the main text, the role of the hamiltonian $H$ is played by the regulators $R_\psi$ and $R_{\psi_c}$, and
 $ is \sim \frac{1}{M^2}$, see eq. \eqref{tra} (here we use a minkowskian  set-up). 
In $D=4$ the $s$-independent term contains $a_2(x,H)$, which produces the anomalies. 

Let us now specialize to the regulator ${R}_\psi   = -  \DDslash^2(A,B) $ which is expanded as
\be 
\begin{aligned}
{R}_\psi   &= -  \DDslash^2(A,B) \\ 
& =  -D^a(A) D_a(A) + B^2 - \gamma^5 (D^a(A) B_a) \\
&-\frac12 \gamma^{ab}
\Big (\hat F_{ab} - 4B_a B_b +\gamma^5( \hat G_{ab} - 4 B_a D_b(A) ) \Big)  
\end{aligned}
\ee
and contains the Bardeen curvatures $\hat F_{ab} $ and $ \hat G_{ab}$ given in \eqref{Bardeen-curvatures}, 
the covariant derivative $D_a(A) =\partial_a+A_a$, and the covariant divergence of $B_a$, 
$D^a(A) B_a = (\partial^a B_a) +[ A^a, B_a]$. 

Comparing it with the heat kernel operator $H$ in eq. \eqref{B1}
\be
H = - \nabla ^2 +V = -\partial^a\partial_a - 2W^a\partial_a -(\partial_a W^a) - W^2 +V \;.
\label{C2}
\ee
allows one to fixes
\begin{align}
W_a &= A_a +\gamma_{ab}\gamma^5 B^b
\\
V &=  -2 B^2 -\gamma^5 (D^a(A) B_a) -\frac12 \gamma^{ab} \hat F_{ab}
\\
  {\cal F}_{ab} &= F_{ab}(A) + (\gamma_{ac}\gamma_{bd}- \gamma_{bc}\gamma_{ad}) B^c B^d 
+ \gamma^5 (\gamma_{ca} D_b(A) B^c -\gamma_{cb} D_a(A) B^c) \;.
\end{align}

Now the coefficient $a_2(R_\psi)$ can be made explicit  using \eqref{SdW}.
We compute directly the relevant Dirac traces, and
list some intermediate results for the reader interested in checking our calculations. 
Recalling the three contributions in the last line of \eqref{SdW},
we find (with $D_a \equiv D_a(A) $):

$i)$ from  $a_2 =\frac12  V^2$ 
\be
\begin{aligned}
\tr [ \gamma^5 T^\alpha a_{2} ({R}_\psi)] 
&= 
\tr_{_{\!\! YM}}  T^\alpha \bigg [ \frac{i}{2} \epsilon^{abcd} \hat F_{ab}   \hat F_{cd}  
+ 4 \{B^2, DB \}  \bigg ]
\\
\tr [T^\alpha a_{2} ({R}_\psi)] 
&=  \tr_{_{\!\! YM}}  T^\alpha \bigg[ -\hat F^{ab}  \hat F_{ab} + 8 B^4  +  2 DB DB\bigg ] 
\\
\tr [ a_{2} ({R}_\psi)] 
&=   \tr_{_{\!\! YM}} \bigg[ -\hat F^{ab}  \hat F_{ab} + 8 B^4  +  2 DB DB\bigg ] 
\end{aligned}
\ee

$ii)$ from  $a_2 =-\frac{1}{6} \nabla^2 V$ 
\be
\begin{aligned}
\tr [ \gamma^5 T^\alpha a_{2} ({R}_\psi)] 
&= 
\tr_{_{\!\! YM}}  T^\alpha \bigg [ -\frac23 i \epsilon^{abcd} \Big (  B_a B_b \hat F_{cd}  + 2 B_a \hat F_{bc}  B_d +\hat F_{ab}  B_c B_d \Big ) 
+\frac23 D^2 D B  
\\
&
+\frac13 [\hat F^{ab},\hat G_{ab}]
+\frac43 [D^a \hat F_{ab}, B^b]  - 2 \{ B^2,  DB \} + 4 B_a DB B^a \bigg ]
\\
\tr [T^\alpha a_{2} ({R}_\psi)]  
&= 
\tr_{_{\!\! YM}}  T^\alpha \bigg[ \frac{i}{6} \epsilon^{abcd} 
\Big ( [ \hat G_{ab}, \hat F_{cd}] -4 [ B_a,  D_b \hat F_{cd}]  \Big )
+8 (B_aB^2 B^a- B^4)   
\\
& +\frac43 D^2 B^2 
+\frac43 (\hat F^{ab} B_a B_b+ 2 B_a\hat F^{ab} B_b+B_a B_b \hat F^{ab} )  \bigg ] 
\\
\tr [ a_{2} ({R}_\psi)] 
&=   \tr_{_{\!\! YM}}  \bigg[ \frac43 D^2 B^2 \bigg ]
\end{aligned}
\ee

$iii)$ from $a_2 = \frac{1}{12} {\cal F}_{ab}^2 $ 
\be
\begin{aligned}
\tr [ \gamma^5 T^\alpha a_{2} ({R}_\psi)] 
&= 
\tr_{_{\!\! YM}}  T^\alpha \bigg [  i \epsilon^{abcd} \Big (\frac16 \hat G_{ab}  \hat G_{cd} 
-\frac23 \{ \hat F_{ab} , B_c B_d\}  +\frac{16 }{3}  B_a B_b  B_c B_d \Big )\\
 &-\frac83 \{ B^2,  DB\} +\frac43  \{ \{ B^a,B^b\}, D_a B_b \}
  \bigg ]
 \\
 \tr [ T^\alpha a_{2} ({R}_\psi)] 
 &=  \tr_{_{\!\! YM}}  T^\alpha \bigg [ 
 \frac13 \hat F^{ab}\hat F_{ab} -\frac43 \{ \hat F_{ab}, B^a B^b \} 
+\frac83( B_a B_b B^a B^b  - B^4) 
 \\
&
 - 8 B_a B^2 B^a  
-\frac43 D_a B_b D^a B^b  -\frac23 DB DB  \bigg ]
 \\
\tr [ a_{2} ({R}_\psi)] 
&=  \tr_{_{\!\! YM}} \bigg [ 
 \frac13 \hat F^{ab}\hat F_{ab} -\frac83 \hat F_{ab}B^a B^b 
   +\frac83 B_a B_b B^a B^b  - \frac{32}{3}B^4 
\\ &
  -\frac43 D_a B_b D^a B^b  -\frac23 DB DB  \bigg ] \;.
 \end{aligned}
 \ee

The analogous results for $a_2(R_{\psi_c})$  are obtained by replacing $A\to -A^T$ and $B\to B^T$ 
(and also $T^\alpha \to T^{\alpha T}  $ for the explicit $T^\alpha$ appearing in the traces). 
Their effect is just to double the contribution from   $a_2(R_{\psi})$ in the chiral and trace anomalies.

\vfill\eject

\end{document}